\documentclass[
    ,final            
  ]
  {aipproc}

\layoutstyle{6x9}

\begin{document}

\def\GeV{{\rm GeV}}
\def\msbar{\overline{MS}}

\title{ Recent Progress in Parton Distributions and Implications 
for LHC Physics}


\author{Robert S. Thorne}{
  address={Cavendish Laboratory, University of Cambridge,
           Madingley Road, Cambridge, CB3 0HE, UK}
}

\iftrue
\author{A.~D. Martin}{address={Institute for Particle Physics Phenomenology,
University of Durham, DH1 3LE, UK},
}

\author{R.~G. Roberts}{
  address={Rutherford Appleton Laboratory, Chilton, Didcot, Oxon, 
OX11 0QX, UK},
}

\author{W.~J. Stirling}{
  address={Institute for Particle Physics Phenomenology,
University of Durham, DH1 3LE, UK},
}
\fi

\begin{abstract}I outline some of the most recent developments in the global 
fit to parton distributions performed by the MRST collaboration.  
\end{abstract}

\maketitle

\begin{figure}[b]
  \includegraphics[height=.35\textheight]{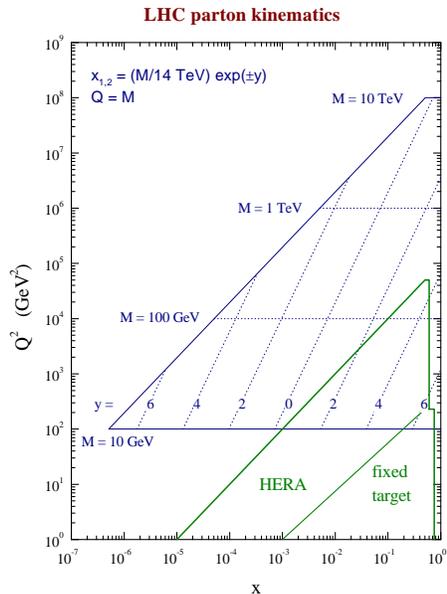}
  \caption{The kinematic range for particle production at the LHC 
}
\end{figure}

At present there is a great deal of interest in the importance of parton 
distributions for studies at the LHC. This necessarily involves a 
certain kinematic range for the partons. The kinematic range for particle 
production at the LHC is shown in fig. 1.
Parton distributions at $x \sim 0.001-0.01$ are  
vital for understanding the standard production 
processes at the LHC. However, even smaller (and higher) $x$ partons are 
required when 
one moves away from zero rapidity, e.g. when calculating the 
total production cross-section of the heavy boson. 
As well as the central values one needs the uncertainties on the partons,
and there has been a lot of work on this \cite{Botje}--\cite{MRSTerror}.
This uncertainty is shown for the $\bar u$ and $\bar d$ quarks 
in fig. 2.  Central rapidity production of $W, Z$ Higgs at the LHC
probes $x=0.006$, which is ideal for the MRST partons.
The current best estimate for the uncertainty due to 
experimental errors is $\delta \sigma_{W,Z}^{\rm NLO}
(\mbox{expt. pdf})  = \pm 2\%$, but we note that there is a theoretical 
uncertainty, which is potentially large due to possible problems at small 
$x$. This is because the large rapidity $W$ and $Z$  
cross-sections sample very small $x$. However, the ratio 
$\sigma(W^+)/\sigma(W^-)$ is a {\it gold-plated} prediction, where
$
R_{\pm } = \frac{\sigma(W^+)}{\sigma(W^-)} \simeq 
\frac{u(x_1)\bar d(x_2)}{d(x_1)\bar u(x_2)}
\simeq \frac{u(x_1)}{d(x_1)}
$
and using the MRST2001E partons
$\delta R_{\pm}(\mbox{expt. pdf})  = \pm 1.4\%$.
Assuming all other uncertainties cancel, this is probably the most 
accurate SM cross-section test at LHC.

\begin{figure}
  \includegraphics[height=.36\textheight]{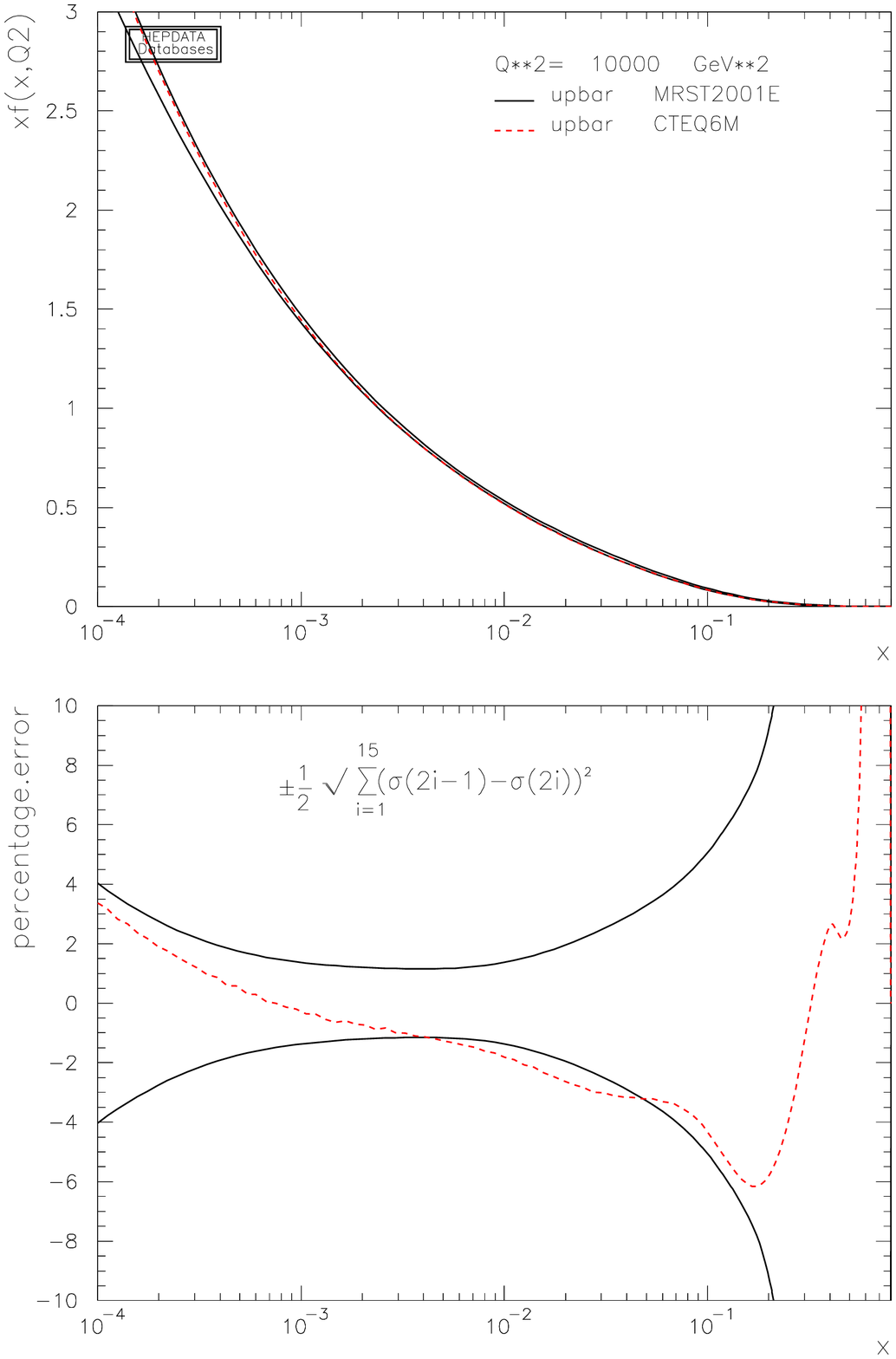}
  \includegraphics[height=.36\textheight]{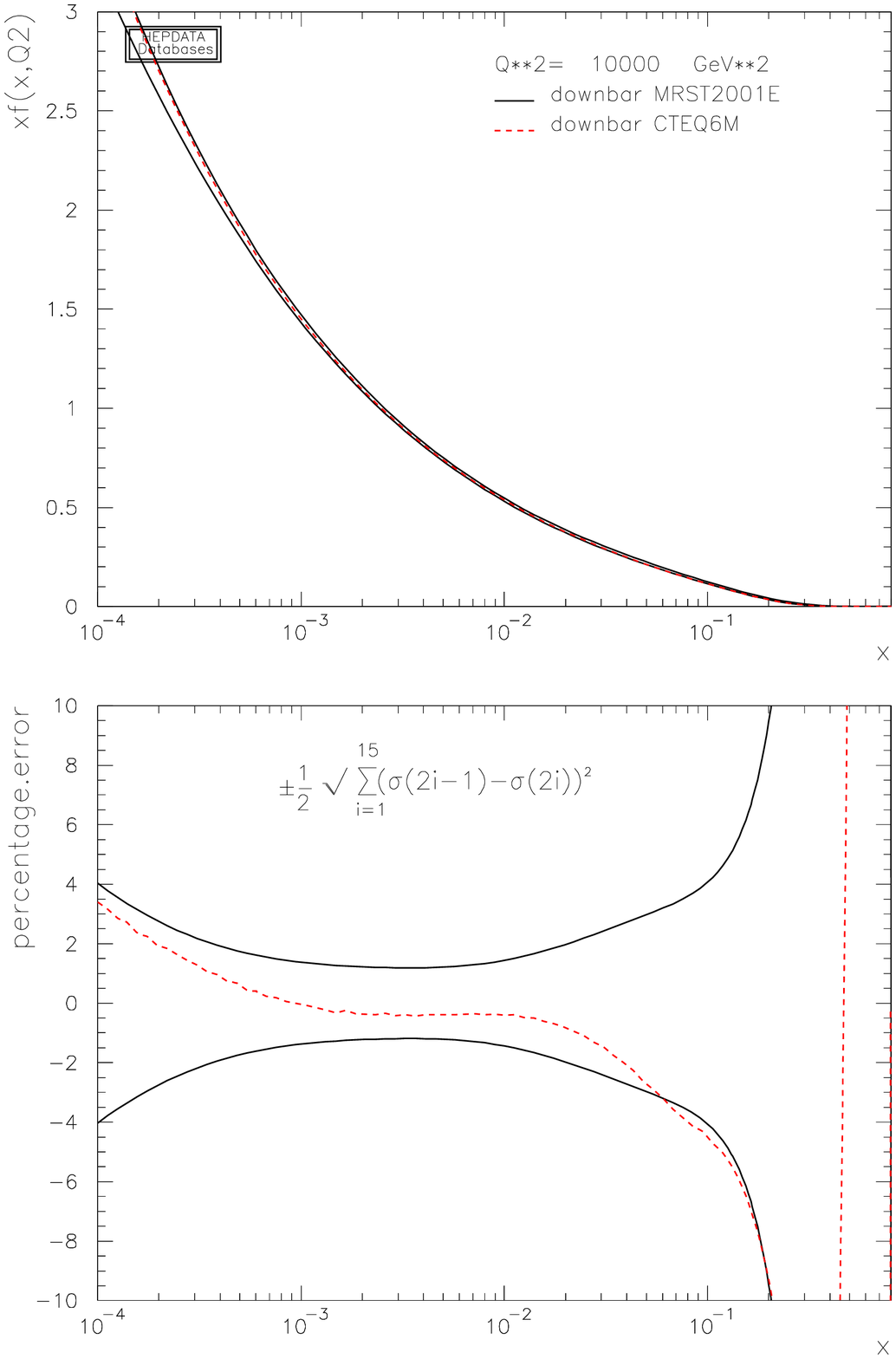}
  \caption{Uncertainty on MRST $\bar u$ and $\bar d$ 
distributions, along with CTEQ6 
\vspace{-0.7cm}}
\end{figure}

This suggests that $\sigma(W)$ or  $\sigma(Z)$ could be used to calibrate 
other cross-sections, e.g. $\sigma(WH)$, $\sigma(Z')$. As an example we 
consider $W$ plus Higgs production.
$\sigma(WH)$ is more precisely predicted than $\sigma(W)$ because it samples 
quark pdfs at higher $x$ and scale.
However, the ratio shows no improvement in uncertainty, and can be 
worse, see fig. 3. 
This is because partons in different regions of $x$ are often 
anti-correlated
rather than correlated, partially due to sum rules.
Similarly, there is no obvious advantage in using $\sigma(t\bar t)$ 
as a calibration SM cross-section, except maybe for very particular, and 
rather large, $M_H$, where the gluon is probed in the same region for both.
However, a light (SM or MSSM) Higgs is dominantly produced via 
$gg\to H$, and the cross-section
has a small pdf uncertainty because $g(x)$ at small $x$ is well 
constrained by HERA DIS data. 
The current best MRST estimate, for $M_H = 120\,\GeV$, 
is $\delta \sigma_{H}^{\rm NLO}(\mbox{expt pdf})  = \pm 2-3\%$
with less sensitivity to small $x$ than $\sigma(W)$.
This is a much smaller  uncertainty than that from higher-order corrections, 
for example \cite{cathiggs}, $\delta \sigma_{H}^{\rm NNLL}
(\mbox{scale variation})  = \pm 8\%$. In constrast, the error on predictions 
for very high-$E_T$
jets at the LHC is dominated by the parton uncertainties, because it is 
sensitive to the relatively poorly known high-$x$ gluon.

\begin{figure}
  \includegraphics[height=.35\textheight]{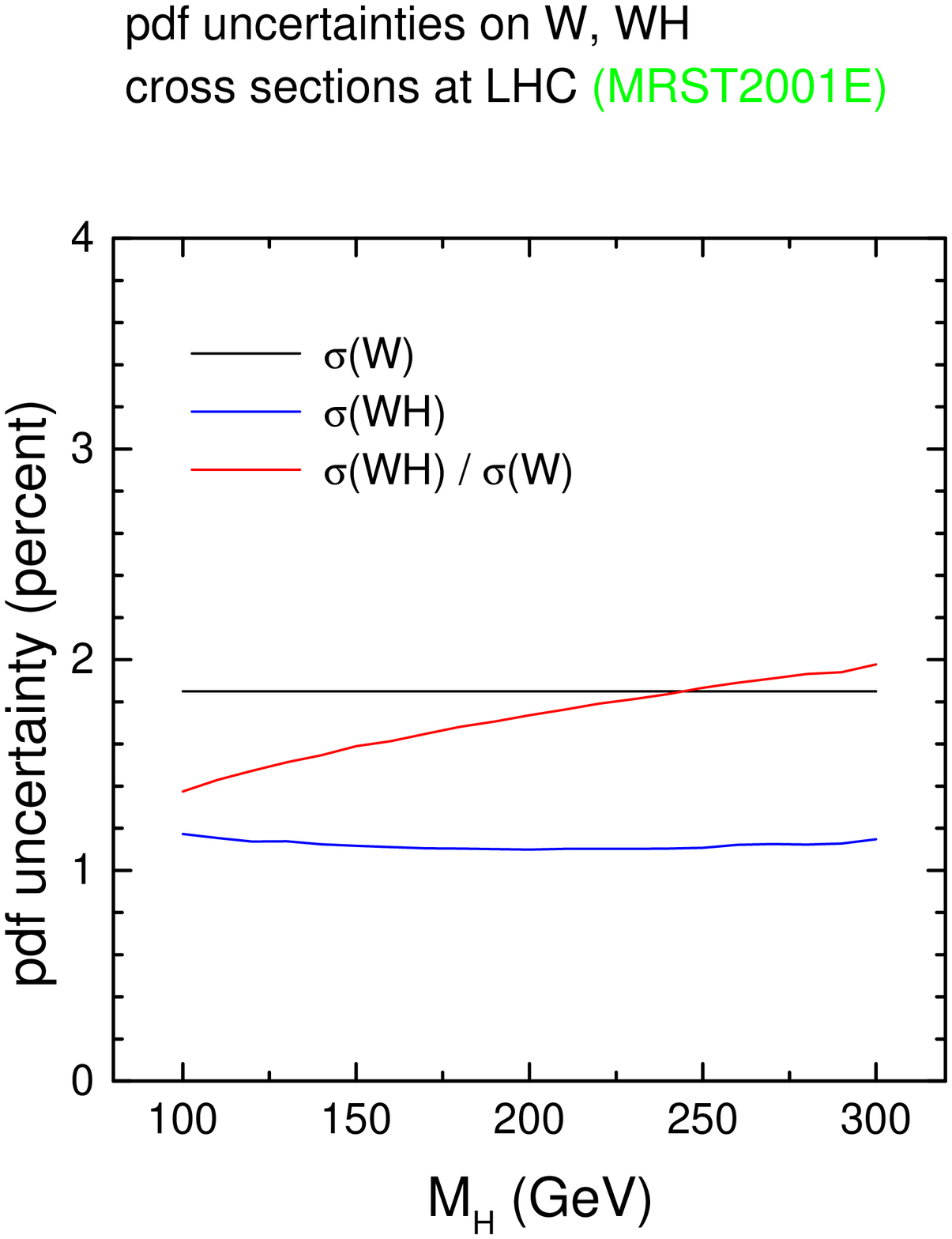}
  \caption{The uncertainty on $W$ and Higgs production 
\vspace{-0.7cm}}
\end{figure}

\begin{figure}[b]
  \includegraphics[height=.38\textheight]{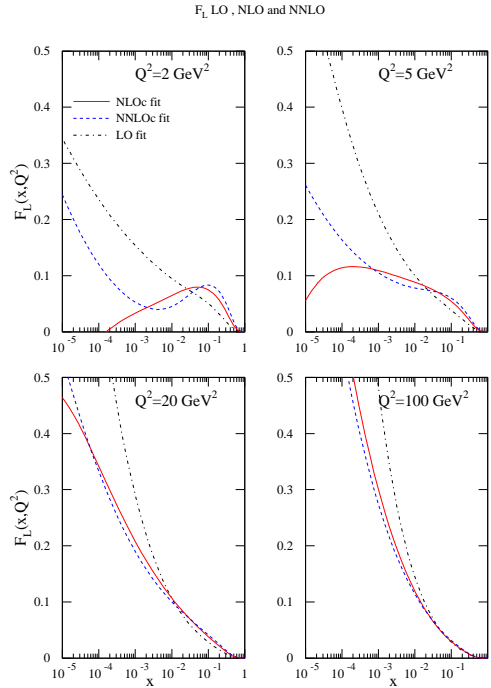}
  \caption{The MRST prediction for $F_L(x,Q^2)$ at LO, NLO and NNLO 
\vspace{-0.7cm}}
\end{figure}

Different approaches to fits generally lead to similar uncertainties 
for measured quantities, but can lead to different central 
values \cite{MRSTerror}. For the true uncertainty one must consider the 
effect of assumptions made during the fit and the correctness of fixed order 
QCD. The failings of NLO QCD are indicated by some areas where 
the fit quality could be improved.
There is a good fit to HERA data, but there are some problems at the highest 
$Q^2$ at moderate $x$, i.e. in $d F_2/d \ln Q^2$. 
Also the data require the gluon to be small or negative at low $Q^2$ and $x$, 
and this is needed by all data (e.g. Tevatron 
jets), not just low-$Q^2$, low-$x$ data. 
Other groups find similar problems with the gluon at low $x$.
CTEQ have a valence-like input gluon at $Q_0^2=1.69\GeV^2$
which would marginally prefer to be negative \cite{newCTEQ}.
There is also instability in the physical, gluon-dominated quantity 
$F_L(x,Q^2)$ going from LO to NLO to NNLO, seen in fig. 4. 
The exact NNLO coefficient function \cite{FLNNLO} has a very large effect, 
a possible sign of $\ln(1/x)$ corrections being required.

As an example of the effect assumptions can make to the fit, MRST found only 
a reasonable fit to jet data \cite{D0,CDF}, but needed to use the 
large systematic errors, while the result is better for CTEQ6 \cite{CTEQ6} 
due to different cuts on other data,
and a different type of high-$x$ parameterization. However, 
for the CTEQ6.1M partons, which give a good fit to the jet data,
the gluon is very
hard as $x \to 1$. MRST have recently utilised the fact that, under a 
change of scheme from $\msbar$ to DIS schemes,
the scheme transformation will dominate the high-$x$ gluon if 
valence quarks are naturally biggest at high $x$ \cite{MRSTdisglu}. 
This allows a
high-$x$ gluon in the $\msbar$ scheme which is determined from the quarks.
At NLO the $\chi^2$ for jets reduces from $154$ to $116$.
This prescription works even better at NNLO -- $\chi^2$ for the jets goes
from $164$ to $117$, and the total $\Delta \chi^2 = -79$.

\begin{figure}
  \includegraphics[height=.38\textheight]{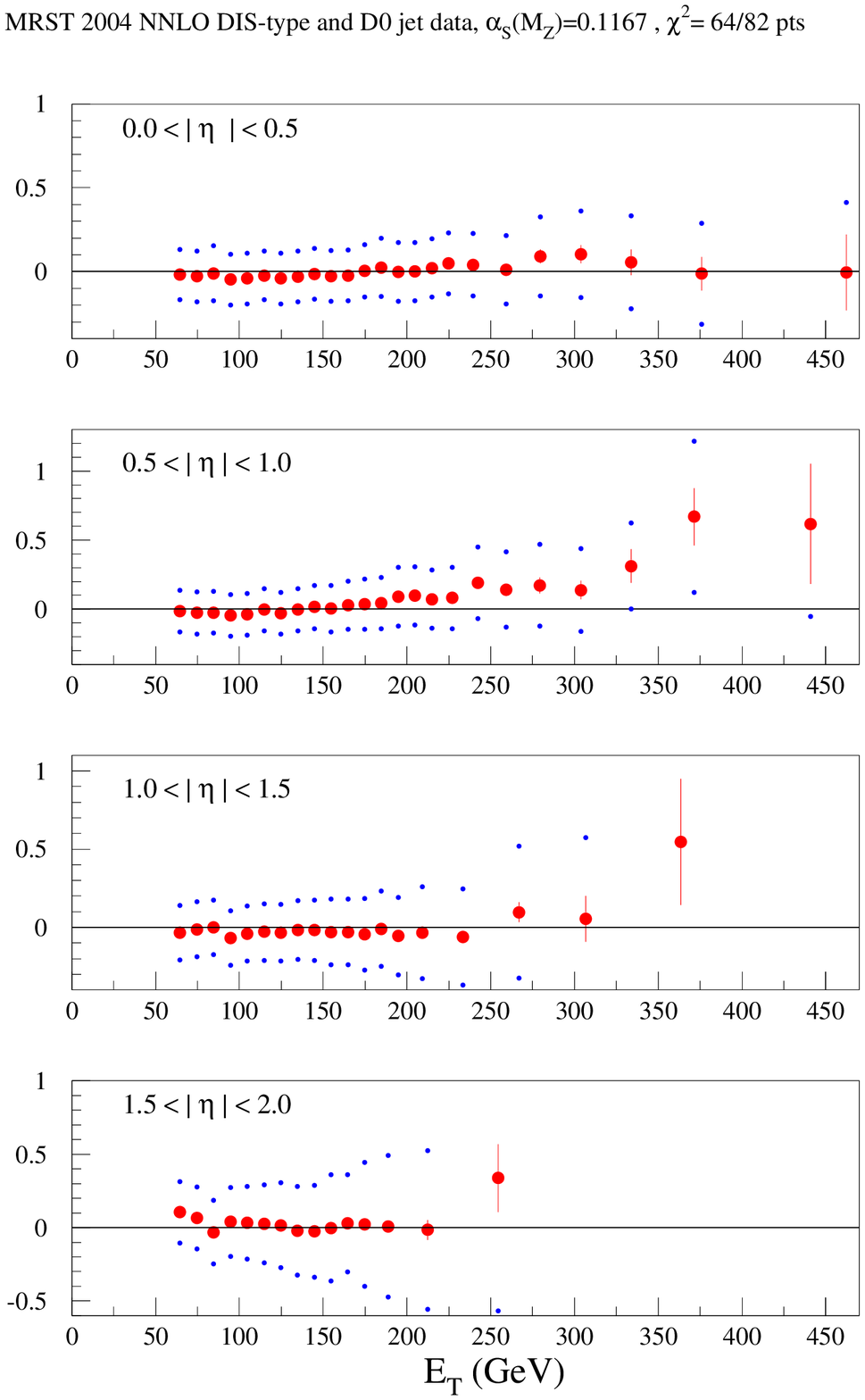}
  \includegraphics[height=.38\textheight]{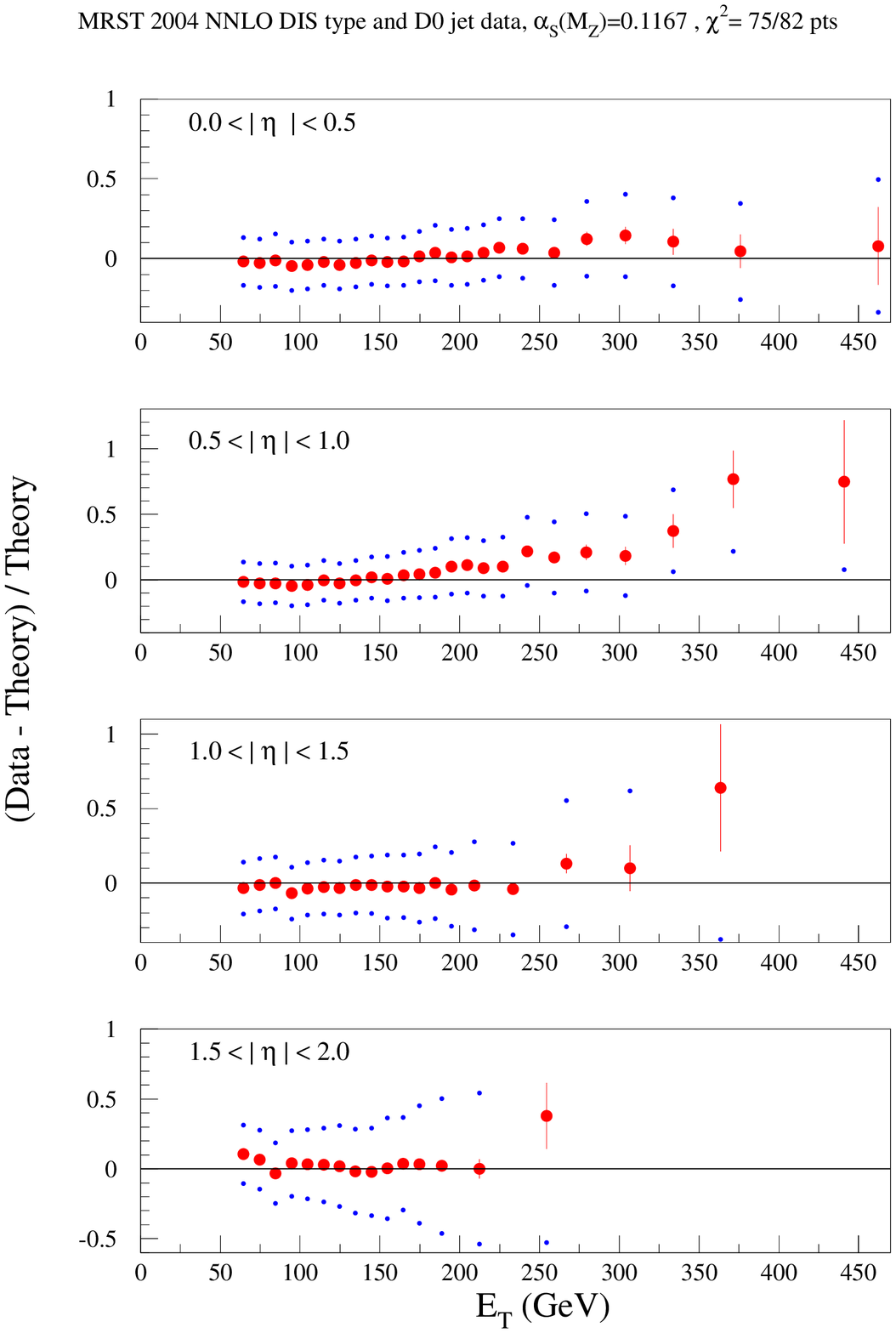}
  \caption{Change in fit to D0 data with weak corrections 
\vspace{-0.7cm}}
\end{figure}

Regarding high $E_T$ jets, there has recently been a calculation of 
weak corrections \cite{Moretti}, which implies that
$\sigma_{QCD} \to \sigma_{QCD}(1
-\frac{2}{3}C_F\frac{\alpha_W}{\pi}\log^2(E_T^2/M_W^2))$.
This is a $12\%$ correction at $E_T =450\GeV$, and the authors
question the validity of recent partons due to this. 
We have studied the phenomenological impact, and 
the movement of both CDF and D0 data is relatively small, 
as shown in fig. 5. The total $\chi^2$ changes by $\sim 15$
without refitting, which is significant but not a disaster. 
The correction is more important at higher $E_T$, but there are 
positive real corrections to be added which depend on the jet 
definitions.

There has also been a study of the inclusion of QED effects by MRST
\cite{MRSTQED}. 
The overall effect is small, but does lead to small isospin 
violation because $u_V^p(x)$ quarks radiate more photons than 
$d_V^n(x)$ quarks. This is in the correct direction to reduce the 
NuTeV $\sin^2 \theta_W$ anomaly \cite{Nutev}, 
and with current quark masses it is halved. Our approach is
supported by data on wide-angle photon scattering,
i.e. $ep \to e \gamma X$ \cite{zeusphot} where the final state electron 
and photon have equal and opposite large transverse momentum.

A  much more important correction is NNLO QCD.
The NNLO splitting functions are now complete \cite{nnlosplit}, 
but very similar 
to the average of previous best estimates, so lead to no large change 
in our previous NNLO partons \cite{MRSTnnlo}. 
The NNLO corrections improve the quality of the fit 
slightly,  and reduce $\alpha_S$. 
However, to perform an absolutely correct NNLO fit we need 
not only exact NNLO splitting functions, but 
also require a rigorous treatment of heavy quark thresholds \cite{thornehf}, 
NNLO Drell-Yan cross-sections \cite{NNLODY}, and a complete treatment of
uncertainties. All this is in hand, and  
an essentially full NNLO determination of partons will appear 
very soon. Only the NNLO jet cross-section is missing. This is
probably not too important -- the  NLO 
corrections themselves are not large, except at high rapidities,
being $\sim 10\%$ at central rapidities. 
There are also good NNLO estimates,
i.e. the threshold correction logarithms, which  are expected to be a 
major component of 
the total NNLO correction \cite{kidonakis}. These give a
flat $3-4\%$ correction, i.e. smaller than the systematic errors on the data. 
Hence, the mistakes from ignoring jets in the fits are bigger than 
the mistakes made at 
NNLO by not knowing the exact hard cross-section.
There is reasonable stability order by order for 
(quark-dominated) $W$ and $Z$ cross-sections, but 
this fairly good convergence is largely guaranteed because 
the quarks are fit directly to the data. 
This is much worse for gluon-dominated quantities,
e.g. $F_L(x,Q^2)$, which is unstable at small $x$ and $Q^2$, 
as seen in fig. 4.

Given this theoretical uncertainty, we devised an approach to look for 
the safe theoretical regions, i.e. change $Q^2_{cut}$ 
and $x_{cut}$, re-fit  and see if the quality of the 
fit to the remaining data improves 
and/or the input parameters change dramatically \cite{MRSTtheory}. 
Raising $Q^2_{cut}$ from $2\GeV^2$ in steps, there 
is a  slow, continuous and significant improvement for higher $Q^2$ up to 
$> 10 \GeV^2$. Raising $x_{cut}$ from $0$ to $0.005$,
there is a continuous improvement, and 
at each step the moderate $x$ gluon becomes more 
positive. This led to the MRST2003 conservative partons, which should be
the most reliable method of parton determination, 
but are {\it only} applicable for a restricted range of $x$ and $Q^2$. 
We also have NNLO conservative partons, with similar cuts and 
improvement in fit quality, but the change in the partons is considerably 
less in this case because NNLO includes important theoretical corrections
lacking at NLO.
The variation in predictions with the cuts indicates the range of possible 
theoretical errors. There is a large change in $\sigma_W$ at the LHC since 
this is sensitive to the 
low $x$ region. The prediction is  much more stable at NNLO, and 
LHC uncertainties are $\sim 3-4 \%$ 
including the theoretical uncertainty. Hence, $\sigma_W$ is a good candidate 
for luminosity determination.
CTEQ have repeated this type of analysis and see a similar type of behaviour 
with cuts \cite{newCTEQ},
although much less dramatic. 
With conservative cuts on data their input gluon again marginally
prefers to have a negative component, confirming that a
negative/small gluon at low $x$ and $Q^2$ is 
not due to the data at low $x$ and $Q^2$. 
They also find that the prediction for $\sigma_W$ at the LHC
moves down, but only a little, as more cuts are imposed. 
However, the loss of data with more cuts leads to larger
errors, and the $\chi^2$ profile is very flat indeed in the 
downwards direction, as seen in fig. 6. 
There is not really any inconsistency with MRST.
If one is cautious about the accuracy of theory at low $x$ 
and $Q^2$, the conclusion that the uncertainty is large on small $x$-sensitive
quantities holds. CTEQ claim no reason to be cautious.  
This theoretical uncertainty is not so much of an issue at NNLO though, as 
discussed above.

\begin{figure}
  \includegraphics[height=.3\textheight]{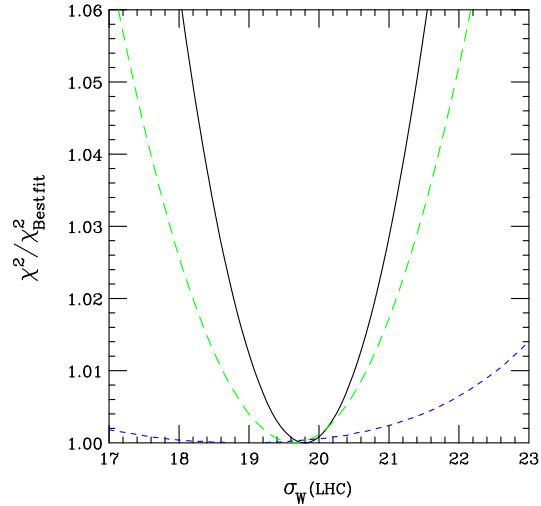}
  \caption{CTEQ $\chi^2$ profile for $\sigma_W$ \cite{newCTEQ}, 
where the wide profile is for
conservative cuts 
\vspace{-0.7cm}}
\end{figure}

In conclusion, we determine the parton distributions and predict
cross-sections by performing global fits,  and the fit quality 
using NLO or NNLO QCD is fairly good. 
There are various ways of looking at uncertainties
due to errors on data, and they are   
$1-5 \%$ for most LHC quantities. Ratios often do not reduce 
uncertainties. QED corrections are small, but introduce important 
isospin asymmetry. 
The uncertainty from using different approaches is 
often comparable to or even larger than deriving from the errors on the data. 
For example, a model for the input form of the gluon
can solve the apparent high-$E_T$ jet problem.
Errors from higher orders/resummation are potentially large. 
Conservative cuts on $x$ and $Q^2$ allow an improved fit to the 
remaining data, and altered partons. CTEQ see some effects from this type of
study, but these are much smaller. NNLO is much more stable than NLO, and more 
theoretically reliable. For MRST 
full NNLO fits are imminent, and should become the new standard.


\vspace{-0.4cm}

\begin{thebibliography}{9}

\bibitem{Botje} 
M.~Botje, {\emph Eur. Phys. J.} {\textbf C14} 285 (2000).

\bibitem{Giele}
W.~T.~Giele, S.~Keller and D.~A.~Kosower, {\tt hep-ph/0104052}.

\bibitem{Alekhin} 
S.~I.~Alekhin, {\emph Phys. Rev.} {\textbf D68} 014002 (2003).

\bibitem{CTEQLag} 
CTEQ Collaboration: D.~Stump {\it et~al.}, {\emph Phys. Rev.} 
{\textbf D65} 014012  (2002).

\bibitem{CTEQHes} 
CTEQ Collaboration:
J.~Pumplin {\it et~al.}, {\emph Phys. Rev.} {\textbf D65} 014013 (2002).

\bibitem{CTEQ6} 
CTEQ Collaboration: J.~Pumplin {\it et~al.}, JHEP 0207:012 (2002).

\bibitem{H1fit} 
H1 Collaboration: C.~Adloff {\it et~al.}, {\emph Eur. Phys. J.} {\textbf C30}
1 (2003).

\bibitem{ZEUSfit} 
ZEUS Collaboration: S.~Chekanov {\it et~al.}, 
{\tt hep-ph/0503274}.

\bibitem{MRSTerror}
 A.~D.~Martin, R.~G.~Roberts, W.~J.~Stirling and R.~S.~Thorne,
{\emph Eur. Phys. J.} {\textbf C28} 455 (2003).

\bibitem{cathiggs} 
S.~Catani,D.~de Florian and M.~Grazzini, JHEP 0307:028 (2003). 

\bibitem{newCTEQ} J.~Huston, J.~Pumplin, D.~Stump and W.~K.~Tung, {\tt
hep-ph/0502080}. 

\bibitem{FLNNLO} S.~Moch, J.~A.~M.~Vermaseren and A.~Vogt, {\emph Phys. Lett.} 
{\textbf B606} 123 (2005); {\tt hep-ph/0504242}.  

\bibitem{D0}D0 Collaboration: B.~Abbott {\it et~al.}, {\emph Phys. Rev. Lett.}
{\textbf 86} 1707 (2001).

\bibitem{CDF}CDF Collaboration: T.~Affolder {\it et~al.}, {\emph Phys. Rev.}
{\textbf D64} 032001 (2001).

\bibitem{MRSTdisglu}  A.~D.~Martin, R.~G.~Roberts, W.~J.~Stirling 
and R.~S.~Thorne,
{\emph Phys. Lett.} {\textbf B604} 61 (2004).

\bibitem{Moretti} S.~Moretti, M.~R.~Nolten and D.~A.~Ross, {\tt
hep-ph/0503152}.  

\bibitem{MRSTQED} A.~D.~Martin, R.~G.~Roberts, W.~J.~Stirling and R.~S.~Thorne,
{\emph Eur. Phys. J.} {\textbf C39} 155 (2005).

\bibitem{Nutev} 
G.~P.~Zeller {\it et al.}, {\emph Phys. Rev. Lett.} {\textbf 88} 091802 (2002).

\bibitem{zeusphot}  
ZEUS collaboration: S.~Chekanov {\it et al.}, {\emph Phys. Lett.} 
{\textbf B595} 86 (2004).

\bibitem{nnlosplit}
  S.~Moch, {\it et al.}, 
{\emph Nucl. Phys.} {\textbf B688} 101 (2004);
{\emph Nucl. Phys.} {\textbf B691} 129 (2004).

\bibitem{MRSTnnlo}
  A.~D.~Martin, R.~G.~Roberts, W.~J.~Stirling and R.~S.~Thorne,
{\emph Phys. Lett.} {\textbf B531} 216 (2002).

\bibitem{thornehf}
R.~S.~Thorne, {\tt hep-ph/0506251}, these proceedings.


\bibitem{NNLODY} C.~Anastasiou, {\it et al},
{\emph Phys. Rev. Lett.}  {\textbf 91}, 182002 (2003);
{\emph Phys. Rev.} {\textbf D69}, 094008 (2004).

\bibitem{kidonakis} 
N.~Kidonakis and J.~F.~Owens, {\emph Phys. Rev.} {\textbf D63} 054019 (2001).

\bibitem{MRSTtheory}  A.~D.~Martin, R.~G.~Roberts, W.~J.~Stirling and 
R.~S.~Thorne,
{\emph Eur. Phys. J.} {\textbf C35} 325 (2004).

\end{thebibliography}
\end{document}